\documentclass[10pt]{article}
\usepackage{latexsym}
\usepackage{amssymb}
\usepackage{amsmath}
\usepackage{amscd}
\usepackage{amsthm}
\usepackage[left=2cm,top=2cm,right=2.5cm,bottom=1.5cm]{geometry}

\usepackage[dvips]{graphicx}
\usepackage{hyperref}

\begin{document}
\begin{center}
\Large{\bf{Bulk Viscous LRS Bianchi-I Universe with Variable $G$\\
 and Decaying $\Lambda$}}
\vspace{10mm}

\normalsize{Anil Kumar Yadav $\footnote{corresponding author}$ ,\; Anirudh Pradhan$^2$ and Ajay Kumar Singh$^3$}\\ \vspace{4mm} 
\normalsize{$^1$Department of Physics, Anand Engineering
College, Keetham, Agra-282 007, India} \\
\vspace{2mm}
\normalsize{$^{1}$E-mail: abanilyadav@yahoo.co.in}\\
\vspace{4mm}
\normalsize{$^{2,\;3}$Department of Mathematics, Hindu Post-graduate College, Zamania 232 331, Ghazipur, India} \\
\vspace{2mm}
\normalsize{$^{2}$ E-mail: pradhan@iucaa.ernet.in, pradhan.anirudh@gmail.com}
\end{center}
\begin{abstract} 
The present study deals with spatially homogeneous and totally anisotropic locally rotationally symmetric 
(LRS) Bianchi type I cosmological model with variable $G$ and $\Lambda$ in presence of 
imperfect fluid. To get the deterministic model of Universe, we assume that the expansion $(\theta)$ in 
the model is proportional to shear $(\sigma)$. This condition leads to $A=\ell B^{n}$, where $A$,\;$B$ are metric 
potential. The cosmological constant $\Lambda$ is found to be decreasing function of 
time and it approaches a small positive value at late time which is supported by recent Supernovae Ia (SN Ia) observations. 
Also it is evident that the distance modulus curve of derived model matches with observations perfectly.\\  
      
\end{abstract}

\section{Introduction}
Recent astronomical observations of type Ia supernovae with redshift parameter $z \leq 1$ (Perlmutter et al. 1997, 1998, 1999; 
Riess et al. 1998, 2004; Garnavich et al. 1998a, 1998b), Wilkison Microwave Anisotropy Probe (WMAP) (Spergel et al. 2003) 
etc. provide evidence that we may live in low mass density Universe i. e. $\Omega\sim 0.3$ (Riess 1986). The predictions of 
observations lead to a convincing belief in modern cosmology that a part of Universe is filled up with dark energy 
$(\Omega\sim 0.7)$, which may be addressed by suitable cosmological constant. 
There are significant observational evidence that the expansion of the Universe is undergoing 
a late time acceleration (Perlmutter et al. 1997, 1998, 1999; Riess et al. 1998, 2004; 
Efstathiou et al. 2002; Spergel et al. 2003; Allen et al. 2004; Sahni and Starobinsky 2000; 
Peebles and Ratra 2003; Padmanabhan 2003; Lima 2004). This, in other words, amounts to 
saying that in the context of Einstein's general theory of relativity some sort of dark energy, 
constant or that varies only slowly with time and space dominates the current composition of cosmos. 
The origin and nature of such an accelerating field poses a completely open question. The 
main conclusion of these observations is that the expansion of the Universe is accelerating. \\

Among many possible alternatives, the simplest and most theoretically appealing possibility for 
dark energy is the energy density stored on the vacuum state of all existing fields in the Universe, 
i.e., $\rho_{v} = \frac{\Lambda}{8\pi G}$, where $\Lambda$ is the cosmological constant. However, 
a constant $\Lambda$ cannot explain the huge difference between the cosmological constant inferred 
from observation and the vacuum energy density resulting from quantum field theories. In an attempt to 
solve this problem, variable $\Lambda$ was introduced such that $\Lambda$ was large in the early 
universe and then decayed with evolution (Dolgov 1983). Since the pioneering work of Dirak (1938), 
who proposed a theory with a time varying gravitational coupling constant $G(t)$, a number of
 cosmological models with 
variable $G$ and $\Lambda$ have been recently studied by several authors
(Arbab 2003; Sistero 1991; Sattar and Vishwakarma 1997; Pradhan and Chakrabarty
2001; Singh et al. 2008). \\

To describe the relativistic theory of viscosity, Eckart (1940) made the first 
attempt. the theories of dissipation in Eckart formulation suffers from serious short-
coming, viz., causality and stability (Hiskock and Lindblom 1985; Hiskock 1986) regardless of the choice of equation of state. 
The problem arises due to first order nature of the theory, since it considers only first order 
deviation from equilibrium. It has been shown that the problems of the relativistic imperfect 
fluid may be resolved by including higher order deviation terms in the transport equation (Hiskock and Salmonson 1991). 
Isreal and Stewart (1970) and Pavon (1991) developed a fully relativistic formulation of the theory 
taking into account second order deviation terms in the theory, which is termed as "transient" or 
"extended" irreversible thermodynamics (EIT). The crucial difference between the standard Eckart and 
the extended Isreal-Stewart transport equations is that the latter is a differential evolution equations, 
while the former is an algebraic relation. In irreversible thermodynamics, the entropy is no longer conserved, 
but grows according to the second law of thermodynamics. Bulk viscosity arises typically in the mixtures 
either of different species or of species but with different energies. The solution of the full causal theory 
are well behaved for all the times. Therefore the best currently available theory for analyzing dissipative processes 
in the Universe is the Full Isreal-Stewart theory (FIS). Several authors (Kremer et al. 2003; Singh and Beesham 2000;
Debnath et al. 2007; Singh and Kale 2009) and recently Yadav (2010, 2011) have obtained cosmological 
models with dissipative effects. Pradhan et al. (2004); Singh (2009); Singh and Kumar (2009) and Bali and Kumawat (2008) 
have studied matter filled imperfect fluid in different physical context.\\

The simplest of anisotropic models are Bianchi type-I homogeneous models whose spatial sections 
are flat but the expansion or contraction rate are direction dependent. For studying the possible 
effects of anisotropy in the early Universe on present day observations many researchers (Huang 1990; 
Chimento et al. 1997; Lima 1996; Lima and Carvalho 1994; Pradhan and Singh 2004; Pradhan and Pandey 2006; 
Saha 2005, 2006a, 2006b) have 
investigated Bianchi type-I models from different point of view. In this paper, we present the exact solution of 
Einstein's field equations with variable $G$ and $\Lambda$ in LRS Bianchi I space-time in presence of imperfect fluid 
as a source of matter. The paper has following structure. In section 2, the metric and field equations are described. 
The section 3 deals with the exact solution of the field equations and physical behaviour of the model. 
The distance modulus curve is described in section 4. At the end we shall summarize the 
findings.\\

\section{The Metric and Field  Equations}
We consider the LRS Bianchi type I metric of the form
\begin{equation}
\label{eq1}
ds^2 = -dt^2 + A^2dx^2 + B^2 \left(dy^2 + dz^2\right),
\end{equation}
where, A and B are functions of t only. This ensures that the model is spatially homogeneous.\\

The energy-momentum tensor $T^{i}_{j}$ for bulk viscous fluid is taken as
\begin{equation}
\label{eq2} T^{i}_{j} = (\rho + p+\Pi)v^{i}v_{j} + (p+\Pi) g^{i}_{j},
\end{equation}
where $p$ is the isotropic pressure; $\rho$ is the energy density of matter; $\Pi$ is the bulk viscous stress; 
$v^{i}=(0,0,0,1)$ is the four velocity vector satisfying the relations
\begin{equation}
\label{eq3} v_{i}v^{i} = -1.
\end{equation}
The bulk viscous stress is given by
\begin{equation}
\label{eq4} \Pi=-\xi\;v^{j}_{;i},
\end{equation}
where $\xi$ is the bulk viscosity coefficient.\\
The Einstein's field equations with cosmological constant may be written as
\begin{equation}
\label{eq5}
R_{j}^i - \frac{1}{2}g_{j}^{i}R - \Lambda g^{j}_{i} = -8\pi G T_{j}^i.
\end{equation}
The Einstein's field equations (\ref{eq5}) for the line-element (\ref{eq1}) 
lead to the following system of equations
\begin{equation}
\label{eq6}
2\frac{B_{44}}{B} + \frac{B_{4}^2}{B^2}  = -8\pi G (p+\Pi) + \Lambda,
\end{equation}
\begin{equation}
\label{eq7}
\frac{A_{44}}{A} + \frac{B_{44}}{B} + \frac{A_{4}B_{4}}{AB} = -8\pi G (p+\Pi) +\Lambda,
\end{equation}
\begin{equation}
\label{eq8}
\frac{B_{4}^2}{B^2} + 2\frac{A_{4}B_{4}}{AB} = 8\pi G \rho + \Lambda.
\end{equation}
Here, and in what follows, sub in-dices 4 in $A$, $B$ and elsewhere indicates
differentiation with respect to $t$.\\
In view of vanishing divergence of Einstein tensor, we get
\begin{equation}
 \label{eq9}
8\pi G\left[\rho_{4} + (\rho + p + \Pi)\left(\frac{A_{4}}{A}+2\frac{B_{4}}{B}\right)\right] + 8\pi \rho G_{4} + \Lambda_{4} = 0.
\end{equation}
Using Eq. (\ref{eq4}), the energy conservation equation (\ref{eq9}) splits into two equations
\begin{equation}
\label{eq10}
\rho_{4} + (\rho +p +\Pi)\left(\frac{A_{4}}{A}+2\frac{B_{4}}{B}\right) = 0,
\end{equation}
and
\begin{equation}
\label{eq11}
8\pi\rho G_{4} +\Lambda_{4} = 0.
\end{equation}
The average scale factor (a) of LRS Bianchi type I model is defined as 
\begin{equation}
 \label{eq12}
a=(AB^{2})^{\frac{1}{3}}.
\end{equation}
The spatial volume (V) is given by
\begin{equation}
\label{eq13}
V = a^{3} = AB^{2}.
\end{equation}
We define the mean Hubble parameter (H) for LRS Bianchi I space-time as
\begin{equation}
\label{eq14}
H = \frac{a_{4}}{a} = \frac{1}{3}\left(\frac{A_{4}}{A}+2\frac{B_{4}}{B}\right).
\end{equation}
The expansion scalar ($\theta$), shear scalar ($\sigma$) and mean anisotropy parameter ($A_{m}$) are defined as
\begin{equation}
\label{eq15}
\theta =3H = \frac{A_{4}}{A}+2\frac{B_{4}}{B},
\end{equation}
\begin{equation}
\label{eq16} 
 \sigma^{2}=\frac{1}{2}\left(\sum_{i=1}^{3} H_{i}^{2}-\frac{1}{3}\theta^{2}\right),
\end{equation}
\begin{equation}
\label{eq17}
A_{m} = \frac{1}{3}\sum_{i=1}^{3}\left(\frac{H_{i}-H}{H}\right)^{2}.
\end{equation}

\section{Solutions of the Field Equations}
The system of eqs. (\ref{eq6})$-$(\ref{eq8}), (\ref{eq10}) and (\ref{eq11}) is employed to obtain 
the cosmological solution. The system of equations is not closed as it has seven unknown ($A$, $B$, $\rho$, $P$ 
$\Pi$, $G$ and $\Lambda$) to be determined from five equations. Therefore, two additional constraint relating 
these parameter are required to obtain explicit solutions of the system.\\
Firstly, we assume that the expansion $(\theta)$ in the model is proportional to the shear $(\sigma)$. 
This condition leads to 
\begin{equation}
\label{eq18}
A = \ell B^{n},
\end{equation}
where $\ell$ and $n$ are constant of integration and positive constant respectively.\\
Following Luis (1985), Johari and Desikan (1994), Singh and Beesham (1999) and recently Singh and Kale (2009), we 
assume the well accepted power law relation between gravitational constant $G$ and scale factor $a$ as
\begin{equation}
 \label{eq19}
G = G_{0}a^{m},
\end{equation}
where $G_{0}$ and $m$ are positive constants.\\
Equations (\ref{eq6}), (\ref{eq7}) and (\ref{eq18}) lead to
\begin{equation}
\label{eq20}
\frac{B_{44}}{B}+(n+1)\frac{B^{2}_{4}}{B^{2}} = 0.
\end{equation}
The solution of equation (\ref{eq20}) is given by
\begin{equation}
\label{eq21}
B = (k_{1}t+k_{0})^{\frac{1}{n+2}},
\end{equation}
where $k_{0}$ and $k_{1}$ are the constants of integration.\\\\
From equations (\ref{eq18}) and (\ref{eq21}), we obtain
\begin{equation}
\label{eq22}
A = \ell(k_{1}t+k_{0})^{\frac{n}{n+2}}.
\end{equation}
The rate of expansion in the direction of $x$, $y$ and $z$ are given by
\begin{equation}
\label{eq23}
H_{x} = \frac{A_{4}}{A} = \frac{nk_{1}}{(n+2)}\frac{1}{(k_{1}t+k_{0})},
\end{equation}
\begin{equation}
\label{eq24}
H_{y} = H_{z} = \frac{k_{1}}{(n+2)}\frac{1}{(k_{1}t+k_{0})}.
\end{equation}
The mean Hubble's parameter $(H)$, expansion scalar $(\theta)$ and shear scalar $(\sigma)$ are given by
\begin{equation}
\label{eq25}
H = \frac{k_{1}}{3(k_{1}t+k_{0})},
\end{equation}
\begin{equation}
\label{eq26}
\theta=\frac{k_{1}}{(k_{1}t+k_{0})},
\end{equation}
\begin{equation}
\label{eq27}
\sigma^{2} = \frac{(n-1)^{2}k_{1}^{2}}{3(n+2)^{2}}\frac{1}{(k_{1}t+k_{0})^{2}}.
\end{equation}
The spatial volume (V), mean anisotropy parameter $(A_{m})$, and average scale factor $(a)$ are found to be
\begin{equation}
\label{eq28}
V = \ell(k_{1}t + k_{0}),
\end{equation}
\begin{equation}
\label{eq29}
A_{m} = \frac{2(n-1)^{2}}{(n+2)^{2}},
\end{equation}
From equations (\ref{eq26}) and (\ref{eq27}), we obtain
\begin{equation}
\label{eq30}
\frac{\sigma}{\theta}=\frac{(n-1)}{\sqrt{3}(n+2)},
\end{equation}
\begin{equation}
\label{eq31}
a = [\ell(k_{1}t+k_{0})]^{\frac{1}{3}}.
\end{equation}
For specification of $\xi$, we assume that the fluid obeys the equation of state of the form
\begin{equation}
\label{eq32}
p = \gamma \rho,
\end{equation}
where $\gamma\;(0\leq \gamma \leq 1)$ is a constant and it is termed as Equation of state parameter (EoS parameter).\\
Differentiating equation (\ref{eq8}), we obtain
\begin{equation}
\label{eq33}
8\pi G \rho_{4} + 8\pi G_{4} \rho +\Lambda_{4} =  \frac{(2n+1)k_{1}^{2}}{(n+2)^{2}(k_{1}t+k_{0})^{2}} \;.
\end{equation}
From equation (\ref{eq11}), (\ref{eq19}), (\ref{eq31}) and (\ref{eq33}), we obtain
\begin{equation}
\label{eq34}
\rho = \frac{3(2n+1)k_{1}}{8\pi G_{0}\ell^{\frac{1}{3}}m(n+2)^{2}(k_{1}t+k_{0})^{\frac{m+3}{3}}} \;.
\end{equation}
Equations (\ref{eq6})$-$(\ref{eq8}), (\ref{eq19}), (\ref{eq21}), (\ref{eq22}), (\ref{eq31}), (\ref{eq32}) and (\ref{eq34}) yield 
exclusive expression for pressure (p), cosmological constant $(\Lambda)$, Gravitational constant $(G)$ and bulk 
viscous stress $(\Pi)$ as follows,
\begin{equation}
\label{eq35}
p = \frac{3\gamma(2n+1)k_{1}}{8\pi G_{0}\ell^{\frac{1}{3}}m(n+2)^{2}(k_{1}t+k_{0})^{\frac{m+3}{3}}},
\end{equation}
\begin{equation}
\label{eq36}
\Lambda = \frac{(2n+1)k_{1}}{(n+2)^{2}}\left[\frac{k_{1}}{(k_{1}t+k_{0})^{2}}-\frac{3\ell^{\frac{m-1}{3}}}
{m(k_{1}t+k_{0})}\right],
\end{equation}
\begin{equation}
\label{eq37}
G=G_{0}[\ell(k_{1}t+k_{0}]^{\frac{m}{3}},
\end{equation}
\begin{equation}
\label{eq38}
\Pi = -\frac{(2n+1)k_{1}}{8\pi G_{0}(n+2)^{2}[\ell(k_{1}t+k_{0})]^{\frac{m}{3}}}\left[\frac{3(1-\gamma)\ell^{\frac{m-1}{3}}}{m(k_{1}t+k_{0})}-
\frac{2k_{1}}{(k_{1}t+k_{0})^{2}}\right].
\end{equation}

We observe that model has singularity at $t=-\frac{k_{0}}{k}$ which can be shifted to $t=0$, 
by choosing $k_{0}=0$. This singularity is of point type as all scale factors vanish at $t=-\frac{k_{0}}{k}$. 
The parameter $\rho$, $p$ and $\Lambda$ start off with extremely large values. 
From (\ref{eq28}), it can be seen that the spatial volume is zero at $t=-\frac{k_{0}}{k}$ and it increases 
with cosmic time. The parameter $H_{x}$, $H_{y}$, $H_{z}$, $H$, $\theta$ and $\sigma^{2}$ diverse at initial singularity. 
These parameters decrease with evolution of Universe and finally drop to zero at late time. 
\textbf{Fig. 1} depicts the variation of gravitational constant $G$ versus time. From (\ref{eq36}), we observe 
that $\Lambda(t)$ is decreasing function of time and $\Lambda > 0$ for all times. 
\textbf{Fig. 2} shows this behaviour of cosmological constant $\Lambda(t)$. Thus the nature 
of $\Lambda$ in our derived model of the Universe is consistent with recent SN Ia observations.\\
\begin{figure}
\begin{center}
\includegraphics[width=4.0in]{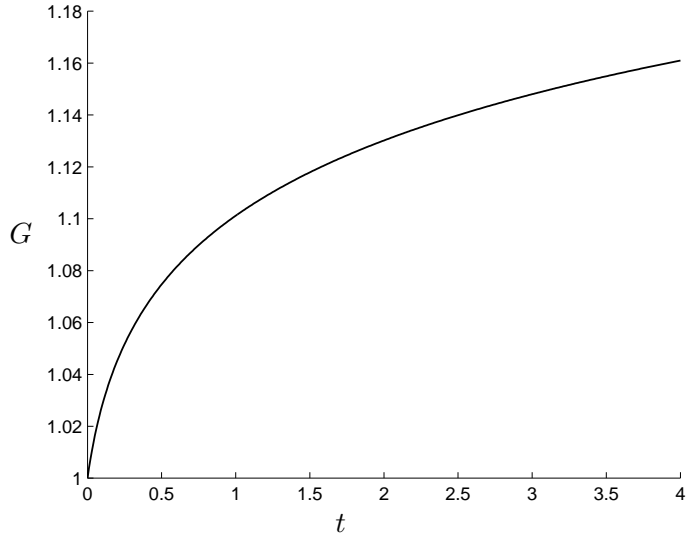} 
\caption{Plot of Gravitational constant $(G)$ versus cosmic time $(t)$.}
\label{fg:ajayF2.eps}
\end{center}
\end{figure}

\begin{figure}
\begin{center}
\includegraphics[width=4.0in]{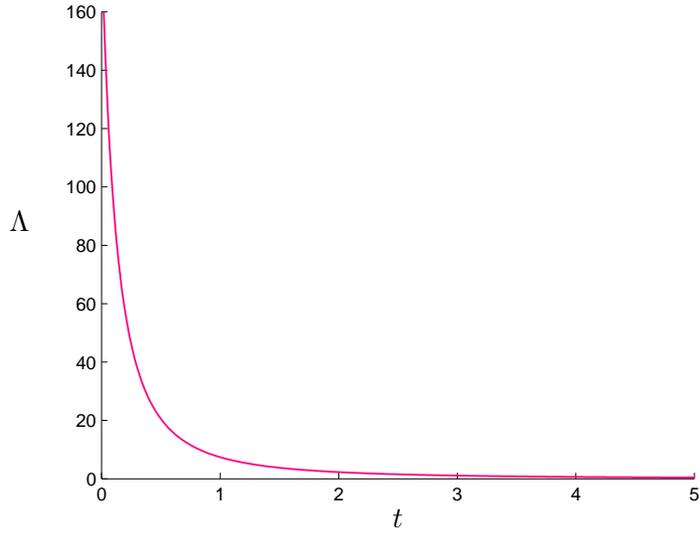} 
\caption{Plot of Cosmological constant $(\Lambda)$ versus cosmic time $(t)$.}
\label{fg:ajayF3.eps}
\end{center}
\end{figure}

\begin{figure}
\begin{center}
\includegraphics[width=4.0in]{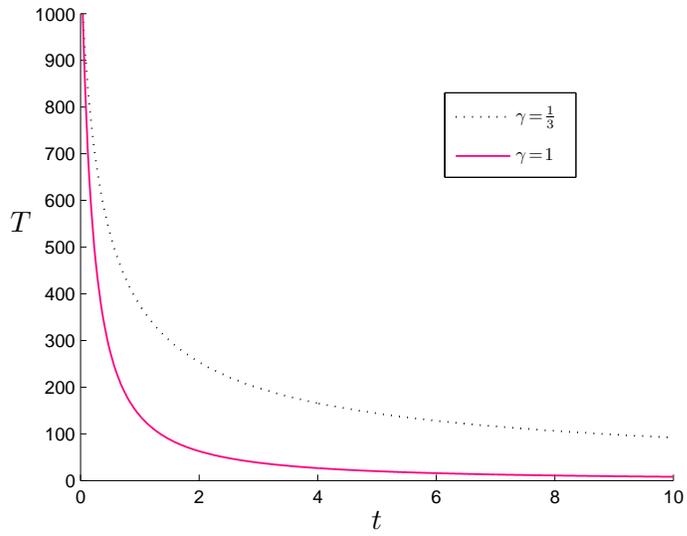} 
\caption{Plot of Temperature $(T)$ versus cosmic time $(t)$ for $\gamma = \frac{1}{3}$ and $\gamma=1$.}
\label{fg:ajayF4.eps}
\end{center}
\end{figure}
 
In EIT, the bulk viscous stress $\Pi$ satisfies a transport equation given by 
\begin{equation}
\label{eq39}
\Pi +\tau \Pi_{4} = -3\xi H - \frac{\epsilon}{2}\tau \Pi\left[3H +\frac{\tau_{4}}{\tau}-\frac{\xi_{4}}{\xi} -\frac{T_{4}}{T}\right]\;,
\end{equation}
where, $\tau$ is the relaxation coefficient of the transient bulk viscous effects and $T\geq 0$ is the absolute temperature 
of the Universe. The parameter $\epsilon$ takes the value $0$ or $1$. Here $\epsilon = 0$, represents truncated Israel-Stewart 
theory and $\epsilon =1$, represents full Isreal-Stewart (FIS) causal theory. One recovers the non-causal Eckart theory for 
$\tau = 0$.\\
Maartens (1995) has pointed out that the Gibb's integrability condition suggest if the equation of state for pressure is 
barotropic (i. e. $p = p(\rho)$) then the equation of state for temperature should be barotropic (i. e. $T = T(\rho)$) and it 
may be expressed as
\begin{equation}
\label{eq40}
T\propto\int{\frac{dp(\rho)}{\rho+p(\rho)}}\;.
\end{equation}
From equations (\ref{eq31}) and (\ref{eq34}), we obtain
\begin{equation}
\label{eq41}
T = T_{0}\rho^{\frac{\gamma}{1+\gamma}}\;,
\end{equation}
where $T_{0}$ stands for a constant.\\
Using Eq.(\ref{eq34}) into Eq. (\ref{eq41}), we obtain the expression for temperature $(T)$ in terms of cosmic time $(t)$ as
\begin{equation}
\label{eq42}
T = T_{0}\left[\frac{3(2n+1)k_{1}}{8\pi G_{0}\ell^{\frac{1}{3}}m(n+2)^{2}(k_{1}t+k_{0})^{\frac{m+3}{3}}}\right]^{\frac{\gamma}{1+\gamma}}\;.
\end{equation}
From equation (\ref{eq42}), it is evident that temperature is decreasing function of time. 
The variation of temperature versus cosmic time for $\gamma = \frac{1}{3}$ (radiation dominated era) and 
$\gamma = 1$ (stiff fluid dominated era) has been graphed in 
\textbf{Fig. 3}. It is clear that temperature of Universe decreases sharply for stiff fluid and 
approaches to small positive value at late time, as expected.\\
\textbf{Bulk Viscosity in Eckart's Theory:} 
The evolution equation (\ref{eq39}) for bulk viscosity in non-causal Eckart's theory reduces to
\begin{equation}
\label{eq43}
\Pi = -3\xi H\;.
\end{equation}
With help of equations (\ref{eq25}), (\ref{eq38}) and (\ref{eq43}), we have the relation 
between bulk viscosity coefficient $(\xi)$ and cosmic time $(t)$ as
\begin{equation}
\label{eq44}
\xi=\frac{(2n+1)}{8\pi G_{0}(n+2)^{2}[\ell(k_{1}t+k_{0})]^{\frac{m}{3}}}\left[\frac{3(1-\gamma)\ell^{\frac{m-1}{3}}}{m}-
\frac{2k_{1}}{(k_{1}t+k_{0})}\right]\;.
\end{equation}
\textbf{Bulk Viscosity in Truncated Theory:}
It has been already pointed out that in truncated theory (i. e. $\epsilon = 0$), the evolution equation (\ref{eq39}) 
for bulk viscosity reduces to 
\begin{equation}
\label{eq45}
\Pi + \tau \Pi_{4} = -3 \xi H\;.
\end{equation}
Following, Singh et al (2009), the relation between $\tau$ and coefficient of bulk viscosity $\xi$ is given by
\begin{equation}
\label{eq46}
\tau = \frac{\xi}{\rho}\;.
\end{equation}
This relation is physically viable because the viscosity signals do not exceed the speed of light. Thus 
the equation (\ref{eq45}) leads to
\begin{equation}
\label{eq47}
\Pi + \frac{\xi}{\rho}\Pi_{4} = -3\xi H\;.
\end{equation}
Using equations (\ref{eq25}), (\ref{eq34}) and (\ref{eq38}) into equation (\ref{eq47}), we obtain
\begin{equation}
\label{eq48}
\xi = \frac{k_{2}\left[3(1-\gamma)\ell^{\frac{m-1}{3}}-2k_{1}(k_{1}t+k_{0})m\right]}{[\ell(k_{1}t+k_{0})^{\frac{m}{3}}]
\left[k_{1}k_{3}(k_{1}t+k_{0})-k_{4}\right]}\;,
\end{equation}
where\\
$k_{2}=\frac{(2n+1)k_{1}}{8\pi G_{0}m(n+2)^{2}}$,\\
$k_{3}= \frac{(m+3)(1-\gamma)+\ell^{\frac{2}{3}}}{\ell^{\frac{2}{3}}}$, \\ $k_{4}=\frac{2m(m+6)k_{1}^{2}}{3\ell^{\frac{m-1}{3}}}$\;.\\
\textbf{Bulk Viscosity in FIS Causal Theory:}
Using equations (\ref{eq35}) and (\ref{eq39}), the transport equation (\ref{eq39}) reduces to
\begin{equation}
\label{eq49}
\Pi + \frac{\xi}{\rho}\Pi_{4} = -3H\xi - \frac{\xi \Pi}{2 \rho}\left[3H - \frac{(1+2\gamma)\rho_{4}}{(1+\gamma)\rho}\right]\;.
\end{equation}
Further, Using equations (\ref{eq25}), (\ref{eq34}) and (\ref{eq38}) into equation (\ref{eq49}), one can easily 
obtain the relation between bulk viscosity coefficient $(\xi)$ and cosmic time $(t)$ as
\begin{equation}
\label{eq50}
\xi = \frac{mk_{1}k_{2}(K_{1}t+k_{2})^{2}\psi(t)}{[\ell(k_{1}t+k_{0})]^{\frac{m}{3}}\left[k_{1}k_{3}(k_{1}t+k_{0})+k_{5}(k_{1}t+k_{0})^{2}\psi(t)-k_{4}\right]}\;,
\end{equation}
where\\
$k_{5}=\frac{m[3(1+\gamma)-(1+2\gamma)(m+3)]}{3(1+\gamma)\ell^{\frac{m-1}{3}}}$\\ 
$\psi(t)=\frac{3(1-\gamma)\ell^{\frac{m-1}{3}}}{m(k_{1}t+k_{0})}-
\frac{2k_{1}}{(k_{1}t+k_{0})^{2}}$\;.\\

\section{Distance Modulus Curves}
The distance modulus is given by
\begin{equation}
\label{eq51}
\mu = 5\;log\;d_{L} +25\;,
\end{equation}
where $d_{L}$ is the luminosity distance and it is defined as
\begin{equation}
\label{eq52}
d_{L} = r_{1}(1+z)a_{0}\;,
\end{equation}
where $z$ and $a_{0}$ represent red shift parameter and 
present scale factor respectively.\\
For determination of $r_{1}$, we assume that a photon emitted by a source with co-ordinate $r=r_{1}$ and $\c{t}=\c{t}_{1}$ 
and received at a time $\c{t}_{0}$ 
by an observer located at $r=0$. Then we determine $r_{1}$ from
\begin{equation}
\label{eq53}
r_{1}=\int_{\c{t}_{1}}^{\c{t}_{0}}\frac{d\c{t}}{a}\;.
\end{equation}
Equation (\ref{eq31}) can be rewritten as
\begin{equation}
\label{eq54}
a=\ell^{\frac{1}{3}}\c{t}^{\frac{1}{3}}\;,
\end{equation}
where $\c{t} = k_{1}t+k_{0}$.\\
Solving equations (\ref{eq51})$-$(\ref{eq54}), one can easily obtain the expression for distance modulus $(\mu)$ 
in term of red shift parameter $(z)$ as
\begin{equation}
\label{eq55}
\mu=5\;log\;\left[\frac{k_{1}}{2H_{0}(1+z)}\left((1+z)^{2}-1\right)\right]+25\;.
\end{equation} 
\vspace{0.5cm}
\begin{center}
 \textbf{Table:\;1}
\end{center}
\begin{tabular}{|c|c|c|c|}
\hline
\;\;\;\;\;\;\;\;\;\;\;Redshift\;$(z)$\;\;\;\;\;\;\;\;\;\;\; & \;\;\;\;\;\;\;\;\;\;\;Supernovae Ia\; $(\mu)$\;\;\;\;\;\;\;\;\;\;\;\;\;\;\; 
& \;\;\;\;\;\;\;\;\;\;\;Our model\; $(\mu)$\;\;\;\;\;\;\;\;\;\;\;\\
\hline
0.014 & 33.73 & 33.81\\
\hline
0.026 & 35.62 & 35.17\\
\hline
0.036 & 36.39 & 35.89\\
\hline
0.040 & 36.38 & 36.13\\
\hline
0.050 & 37.08 & 36.63\\
\hline
0.063 & 37.67 & 37.14\\
\hline
0.079 & 37.94 & 37.66\\
\hline
0.088 & 38.07 & 37.90\\
\hline
0.101 & 38.73 & 38.22\\
\hline
0.160 & 39.08 & 39.29\\
\hline
0.240 & 40.68 & 40.26\\
\hline
0.380 & 42.02 & 41.40\\
\hline
0.430 & 42.33 & 41.71\\
\hline
0.480 & 42.37 & 42.01\\
\hline
0.620 & 43.11 & 42.67\\
\hline
0.740 & 43.35 & 43.15\\
\hline
0.778 & 43.81 & 43.28\\
\hline
0.828 & 43.59 & 43.46\\
\hline
0.886 & 43.91 & 43.64\\
\hline
0.910 & 44.44 & 43.72\\
\hline
0.930 & 44.61 & 43.78\\
\hline
0.949 & 43.99 & 43.83\\
\hline
0.970 & 44.13 & 43.89\\
\hline
0.983 & 44.13 & 43.93\\
\hline
1.056 & 44.25 & 44.13\\
\hline
1.190 & 44.19 & 44.47\\
\hline
1.305 & 44.51 & 44.73\\
\hline
1.340 & 44.92 & 44.81\\
\hline
1.551 & 45.07 & 45.235\\
\hline 
\end{tabular} \\

\vspace{0.5cm}
\indent The observed value of distance modulus $(\mu(z))$ at different redshift parameters $(z)$ given in table 1 (SN Ia Data) 
are employed to draw the curve corresponding to the calculate value of $\mu(z)$. \textbf{Fig. 4} shows the plot of 
observed $\mu(z)$ (dotted line) and calculated $\mu(z)$ (solid line) versus redshift parameters $(z)$.
\begin{figure}
\begin{center}
\includegraphics[width=4.0in]{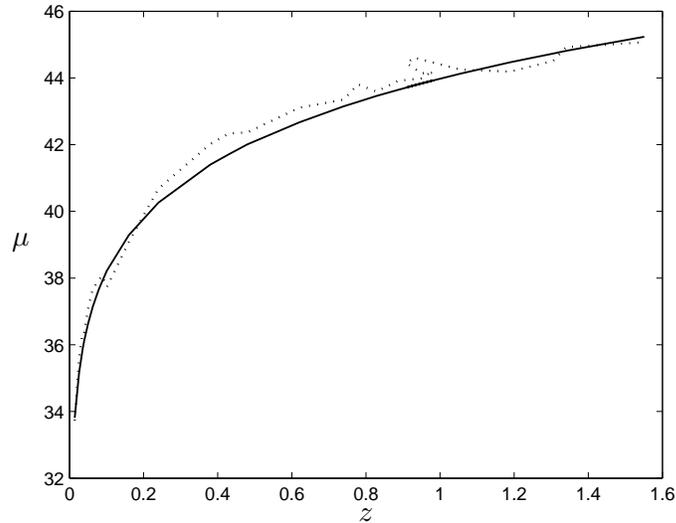} 
\caption{Plot of distance modulus $(\mu)$ versus redshift $(z)$ for Supernova data (dotted line) and for our model (solid line).}
\label{fg:ajayF1.eps}
\end{center}
\end{figure}
\section{Concluding Remarks}
In this paper, we have presented exact solution of Einstein's field equations 
with variable $G$ and $\Lambda$ in LRS Bianchi type I space-time in presence of 
imperfect fluid. The main features of the work are as follows:\\

\begin{itemize}
\item The derived model represents the power law solution which is different 
from other author's solution. It seems to describe the dynamics of Universe from 
big bang to present epoch.\\
\item The cosmological constant $(\Lambda)$ is found to be decreasing function of 
time and it approaches to small positive value at late time. A positive value of $\Lambda$ 
corresponds to negative effective mass density (repulsion). Hence we expect that in the 
Universe with the positive value of $\Lambda$, the expansion will tends to accelerate. Thus 
the derived model predicts accelerating Universe at present epoch. This is in the favour of 
recent supernovae Ia observations.\\
\item The temperature of Universe in derived model is infinitely high at early stage of evolution 
of Universe but it approaches to small positive value at later stage. This means that temperature 
is also decreasing function of time. The same is predicted by CMBR observations.\\
\item If we choose $n=1$, the mean anisotropy parameter 
vanishes. Therefore isotropy is achieved in the derived model for $n=1$. Also we see that 
for $\ell = 1$ and $n=1$, the directional scale factors vary as 
$A(t) = B(t) = a(t)$, therefore metric (\ref{eq1}) reduces to the flat FRW space-time. 
Thus $\ell =1$ and $n=1$, turn out to be the condition of flatness in the derived model. 
It is important to note here that for $n=1$, shear scalar vanishes but the bulk viscosity contributes
to the expansion of Universe and for positive value of $n$, the bulk viscosity coefficient $(\xi)$ decreases with time.\\
\item The distance modulus curve of derived model is in good agreement with SN Ia data (see \textbf{Fig. 4} and Table 1).\\
\item The age of Universe is given by
$$T_{0}=\frac{1}{3}H_{0}^{-1}-\frac{k_{0}}{k_{1}}$$
\end{itemize}
Finally, the model presented in this paper is accelerating, shearing and starts expanding with big bang singularity. 
This singularity is of point type singularity.\\ 
\section*{Acknowledgements} 
The authors would like to thank the anonymous referee for his/her valuable comments which improved the paper in this form. 
One of the authors (A. K. Yadav) would like to thank The Institute of Mathematical Science (IMSc), Chennai, India for 
providing facility and support where part of this work was carried out. Also the partial support by the State Council of 
Science and Technology, Uttar Pradesh (U.P.), India is gratefully acknowledged by A. Pradhan. 
  

\end{document}